\documentclass[preprint,aps,amsmath,amssymb,letterpaper,onecolumn]{revtex4}
\pdfoutput=1
\usepackage{hyperref}
\usepackage{graphicx}

\def\cal#1{\mathcal{#1}}
\def\av#1{\langle #1 \rangle}

\def\eqq#1{Eq.~(\ref{#1})}
\def\eq#1{(\ref{#1})}

\def\k{{\kappa}}

\def\es{\boldsymbol{S}}
\def\t{\boldsymbol{\tau}}

\def\beq{\begin{equation}}
\def\eeq{\end{equation}}
\def\bea{\begin{eqnarray}}
\def\eea{\end{eqnarray}}

\def\kt{T}

\begin{document}
\title{Nonclassical assembly pathways of anisotropic particles}
\author{Stephen Whitelam\footnote{\texttt{swhitelam@lbl.gov}}} 
\affiliation{Molecular Foundry, Lawrence Berkeley National Laboratory, 1 Cyclotron Road, Berkeley, CA 94720, USA}
\begin{abstract}
Advances in synthetic methods have spawned an array of nanoparticles and bio-inspired molecules of diverse shapes and interaction geometries. Recent experiments indicate that such anisotropic particles exhibit a variety of `nonclassical' self-assembly pathways, forming ordered assemblies via intermediates that do not share the architecture of the bulk material. Here we apply mean field theory to a prototypical model of interacting anisotropic particles, and find a clear thermodynamic impetus for nonclassical ordering in certain regimes of parameter space. In other parameter regimes, by contrast, assembly pathways are selected by dynamics. This approach suggests a means of predicting when anisotropic particles might assemble in a manner more complicated than that assumed by classical nucleation theory.
\end{abstract}
\maketitle

\section{Introduction}
Classical nucleation theory assumes the formation of ordered structures from similarly ordered nuclei~\cite{gibbs1928collected,becker1935kinetic}. Mounting evidence, however, suggests that many molecular and nanoscale systems form ordered structures in more complicated ways, first associating as metastable, often amorphous aggregates. Such nonclassical crystallization~\cite{vekilov2005two,erdemir-nucleation,basios2009new} has been observed in systems of spherical colloids~\cite{zhang2007does,zhang2009nucleation,savage2009experimental} and the globular protein lysozyme~\cite{galkin2000control,filobelo2005ssc,pan2005nucleation}, as well as in numerous simulation studies~\cite{vanmeel2008tsv,wolde1999homogeneous}. 

Computational and theoretical work~\cite{asherie1996phase,wolde1997epc,lutsko2006ted} reveals one set of circumstances in which particles bearing isotropic interactions assembly nonclassically: when attractions are made sufficiently short-ranged, the system's liquid-vapor critical point is submerged (in a density--temperature phase diagram) within the regime of solid-fluid coexistence. In what appears to be an immediate kinetic consequence of this thermodynamics, randomly dispersed components possessing short-ranged isotropic attractions, cooled below the liquid-vapor critical temperature, tend to assemble into ordered solids only after forming transient liquid-like phases. However, most real components, from proteins to ions~\cite{de2003principles} to the plethora of recently-synthesized nanoparticles~\cite{glotzer2004materials}, interact via anisotropic or `patchy' attractions. Simulation work~\cite{rein1998coil,gee2005ass,doye2007controlling, auer2008self,duff2009nucleation} reveals assembly pathways of such components to be in general richer than those of their isotropic counterparts. Further, experiments indicate that anisotropic proteins can crystallize via a metastable dense phase {\em outside} the liquid-vapor coexistence regime~\cite{galkin2000nucleation}, an observation bolstered by recent simulations~\cite{liu2009self}. 

Two important ideas underpin our understanding of nonclassical assembly. The step rule of Ostwald~\cite{ostwald1897studien} states that metastable precursors of the stable phase may appear if those precursors are closer in free energy to the parent phase than is the stable solid. The conjecture of Stranski and Totomanow (ST)~\cite{st} is the closely-related statement that the precursors that emerge are those confronted by the smallest free energy barriers to their nucleation. While these ideas receive broad support~\cite{wolde1999homogeneous}, recent evidence suggests that dynamical effects can invalidate the ST conjecture~\cite{sanz2007evidence}. For all but one-component isotropic particles, then, it seems that there exists no simple physical picture that predicts {\em when} particles might assemble in a nonclassical fashion.

Here we propose a step in this direction by considering a microscopic model prototypical of a collection of particles bearing isotropic and anisotropic interactions. In Sections~\ref{sec_model} and~\ref{sec_phase} we introduce this model and use mean field theory to determine its phase behavior. We summarize this behavior in Fig.~\ref{fig1}. In Section~\ref{sec_order} we focus on thermodynamic states at which the solid phase is stable. We ask how the solid emerges if one begins with a well-mixed system and considers Langevin evolution in a free energy space of bulk `density' and `structure' order parameters. We find that under some conditions there exists a free energetic driving force for assembly of the ordered solid phase via nonclassical pathways. In such cases the free energy surface local to the homogeneous fluid phase is stable in one `direction' of order parameter space, and unstable in the other direction. Consequently, density and structure order parameters evolve sequentially, rather than simultaneously. There also exist thermodynamic states at which no such bias exists. In such cases, assembly pathways are determined principally by order parameter dynamics. We summarize these observations in Fig.~\ref{fig2}. We conclude, in Section~\ref{sec_extension}, by discussing an extension of this model in which the assembly of a solid phase is induced by the formation of a solid intermediate. This discussion is summarized in Fig.~\ref{fig3}.

\section{Model}
\label{sec_model}
We consider a collection of particles that live on the sites $i \in \{1,...,N\}$ of a $d$-dimensional hypercubic lattice. The presence or absence of a particle at site $i$ is signaled by the occupancy variable $n_i$ taking the value 1 or 0, respectively. Particles bear unit orientation vectors $\es_i$, which, for simplicity, we assume to rotate in a plane~\footnote{For Heisenberg spins the Bessel function in \eqq{eq1} is replaced by $2 \, {\rm sinc}(\beta Q \tau)$. The resulting mean field phase behavior is similar to that of the XY-like interaction described in the main text, following the replacements $Q \to \frac{3}{2} Q$ and $\mu \to \mu + T \ln 2$. The {\em spatial} behavior of these two models, however, and the nature of their topological defects is known to be qualitatively different~\cite{lau1989numerical}.}. We impose an energy function $\cal{H}=\sum_{i=1}^N \left(\frac{1}{2z} \sum_j U_{ij}-\tilde{\mu} n_i\right)$, where $j$ runs over the $z=2d$ nearest neighbors of $i$, and $\tilde{\mu}$ is a chemical potential. The dimensionality $d$ profoundly affects the nature of fluctuations within the model, but at mean field level serves only to scale the strength of its pairwise interaction. We choose the pairwise interaction $U_{ij}$ to be a minimal representation of particles able to interact both isotropically and anisotropically:
\beq
\label{eq0}
U_{ij}= - n_i n_j \left(J+Q \es_i \cdot \es_j\right).
\eeq
Here $J$ is the strength of the isotropic interaction, and $Q$ is the strength of the anisotropic interaction. This model is designed to describe vapor- and liquid-like phases of small and large occupancy number, respectively, in which particle orientations $\es_i$ are disordered, and a ferromagnetic solid-like phase of large occupancy number in which particle orientations show a high degree of order (a related coupled Ising-Heisenberg model possessing particle-vacancy symmetry was studied in~\cite{huang1986spin}). We next derive the free energy of this model in a mean field approximation. In such an approximation (see e.g.~\cite{geng2009theory}) the fluctuating variables at a given site feel only the thermal averages of variables at neighboring sites. The effective field at a given site is to this approximation $\cal{H}_{\rm eff}  = -n \left(J \rho+Q  \es \cdot \t  +\tilde{\mu} \right) \equiv U_{\rm eff} -\tilde{\mu} n$. Here $n$ and $\es$ are fluctuating variables, and we have introduced the collective density- and structure order parameters $\rho \equiv \av{n}$ and $\t \equiv \av{n \es}$, respectively. These order parameters serve to distinguish phases of low and high density, and phases in which particle orientations are disordered or mutually aligned. For future notational convenience we also introduce the Ising-like density variable $\phi\equiv 2 \rho-1$; we will use both $\phi$ and $\rho$. Thermal averages are defined self-consistently through the relation $\av{A} \equiv {\rm Tr} \left(A \,P_{\rm eq}\right)$, where the equilibrium measure $P_{\rm eq} = q^{-1}e^{-\beta \cal{H}_{\rm eff}}$ with $q \equiv {\rm Tr}\, e^{-\beta \cal{H}_{\rm eff}}=1+ 2 \pi e^{\beta (J  \rho+\tilde{\mu})} {\rm I}_0(\beta Q  |\t| )$. Here ${\rm I}_n$ is the $n^{\rm th}$ order modified Bessel function of the first kind; $\beta \equiv 1/T$ (we adopt units such that $k_{\rm B}=1$); and the trace ${\rm Tr} (\cdot) \equiv \sum_{n=0,1} \left\{ \delta_{n,1} \int d \es + \delta_{n,0} \right\} (\cdot )$ has been carried out by aligning $\t$ with $\hat{\boldsymbol{e}}_x$. The effective Helmholtz free energy per site is then $f_{\rm eff}(\rho,\tau) = E-TS$, where $E=\frac{1}{2} \av{U_{\rm eff}}-\tilde{\mu} \rho$ and $-TS= \kt \av{\ln P_{\rm eq}}=-\av{\cal{H}_{\rm eff}}-\kt \ln q$. Thus $f_{\rm eff}(\rho,\tau)=-\frac{1}{2} \av{U_{\rm eff}}-\kt \ln q$, or
\bea
\label{eq1}
f_{\rm eff}(\rho,\tau)&=&\frac{1}{2} \left(J \rho^2+Q \tau^2\right) \nonumber \\
&-&\kt \ln \left[ 1+ e^{\beta (J  \rho+\mu)} {\rm I}_0(\beta Q  \tau )\right],
\eea
where $\tau \equiv |\t|$ and $\mu\equiv \tilde{\mu} + T \ln 2 \pi$. We consider \eqq{eq1} to have been divided through by dimensions of temperature, and all parameters in that equation to have been de-dimensionalized accordingly. Equations of state for the density and structure order parameters can be obtained by minimizing the free energy, and read
\beq
\label{eq2}
\rho = \frac{ {\rm I}_0 (\beta Q \tau)}{e^{-\beta(J \rho+\mu)}+{\rm I}_0 (\beta Q \tau)}, 
\eeq
and
\beq
\label{eq3}
\t = \hat{\boldsymbol{e}}_x \frac{  {\rm I}_1(\beta Q \tau)}{ e^{-\beta(J \rho+\mu)}+{\rm I}_0(\beta Q \tau)}.
\eeq
The expressions (\ref{eq1})--(\ref{eq3}) describe phases of vapor (low density, orientationally disordered: $\phi < 0, \tau=0)$, liquid (high density, orientationally disordered: $\phi > 0, \tau=0)$ and solid (high density, orientationally ordered: $\phi > 0, \tau >0)$. In the following section we derive the phase diagrams shown in Fig.~\ref{fig1}. Readers not interested in the details of these calculations should focus on Section~\ref{sec_order}, in which we ask how the solid phase emerges if it is stable and if we start from conditions of moderate density without orientational order.  

\section{Model phase behavior}
\label{sec_phase}
We first focus on the phase behavior of the model when either the isotropic interaction or the anisotropic interaction vanishes. For $Q=0$ we recover from \eq{eq1} -- ignoring field-independent terms and introducing $K\equiv J/4$, $\mu_{\rm coex} \equiv -2 K $ and $h \equiv \frac{1}{2}\left(\mu -\mu_{\rm coex}\right)$ -- the Ising model free energy $f_{\rm I}(\phi) = \frac{K}{2} \phi^2 - \kt \, \ln \cosh \left[ \beta \left(K \phi+h\right) \right]$. We recover from \eq{eq2} the equation of state $\phi = \tanh\left[ \beta(K \phi+h)\right]$. These expressions caricature the thermodynamics of the liquid-vapor phase transition~\cite{binney1992theory}. For $K=0$, Eqs. \eq{eq1}--\eq{eq3} describe, at $\mu=\mu_{\rm coex}$, a continuous phase transition in $\k \equiv Q/4$ from a fluid phase having $\tau=0=\phi$ to a solid phase whose order parameter scales near the critical point $\k_{\rm crit} = \beta^{-1}$ as $\tau_{\rm sol} \sim \left( \k-\k_{\rm crit} \right)^{1/4}$. 

The phase diagram for general values of  $K$ and $\kappa$ (for $T=1$)  is shown in Fig.~\ref{fig1}(a) (henceforth we focus on the case $\mu=\mu_{\rm coex}$). It identifies a homogeneous fluid phase H ($\phi=0=\tau$); a regime of phase-separated (PS) liquid L ($\phi>0 ,\tau=0$) and vapor V ($\phi<0, \tau=0$); and a solid phase S ($\phi>0, \tau>0$). The solid phase is described by \eqq{eq3} with $\rho=\rho_{\rm sol}(\tau)= \tau {\rm I}_0(4\beta \kappa \tau)/{\rm I}_1(4 \beta \kappa \tau)$. The points $(K,\kappa)=(1,0)$ and $(0,1)$ are continuous critical points; C$_1$ and C$_2$ are lines of continuous critical points; and F (which abuts C$_2$) is a line of first order phase transitions. The line M delimits the limit of fluid metastability. The equation of the union of the lines M and C$_2$ is $2 K = \left(\beta - 1/\kappa \right)^{-1} \ln \left( 2 \beta \kappa -1\right)$. It was found by equating derivatives with respect to $\tau$, at $\tau=0$, of each side of \eqq{eq3} (with $\rho=\rho_{\rm sol}(\tau)$). 

Panels (b) and (c) of Fig.~\ref{fig1} show phase diagrams in the density-temperature plane for two choices of $K$ and $\k$. Panel (b) describes a case $(K=1.5,\kappa=0.6)$ in which the solid phase becomes stable only well below the liquid-vapor critical point. Expansion about $\tau=0$ of \eq{eq1} with $\rho=\rho_{\rm sol}(\tau)$ reveals the onset of $\tau$ to be continuous with temperature (see inset), scaling below the solid phase critical temperature $T_{\rm c} = 1.08$ (obtained from $\beta_{\rm c} \k \left(1 + \tanh\left[K\left(\k^{-1}- \beta_{\rm c}\right)\right]\right)=1$) as $\tau_{\rm sol} \sim \left(T_{\rm c} - T \right)^{1/2}$. The density of the solid phase at the critical point is $\rho_{\rm sol}(\tau \to 0)= \left( 2 \k \beta_{\rm c} \right)^{-1} \approx 0.90$. A different scenario is seen in Fig~\ref{fig1}(c): here the solid phase becomes viable above the liquid-vapor critical point (and stable with respect to the homogeneous fluid phase below $T \approx 1.1$) and the onset of $\tau$ is now first order with $\kappa$ (see inset). Cases (b) and (c) loosely resemble phase diagrams of Lennard-Jones particles, with distinct vapor, liquid and solid phases; away from $\mu=\mu_{\rm coex}$ (not shown) the phenomenology of this model is more akin to that of isotropic potentials of shorter range~\cite{wolde1997epc}, where only one fluid phase is stable.

\section{Pathways of assembly of the solid phase} 
\label{sec_order}

\begin{figure*}[b] 
\label{}
\centering
\includegraphics[width=\linewidth]{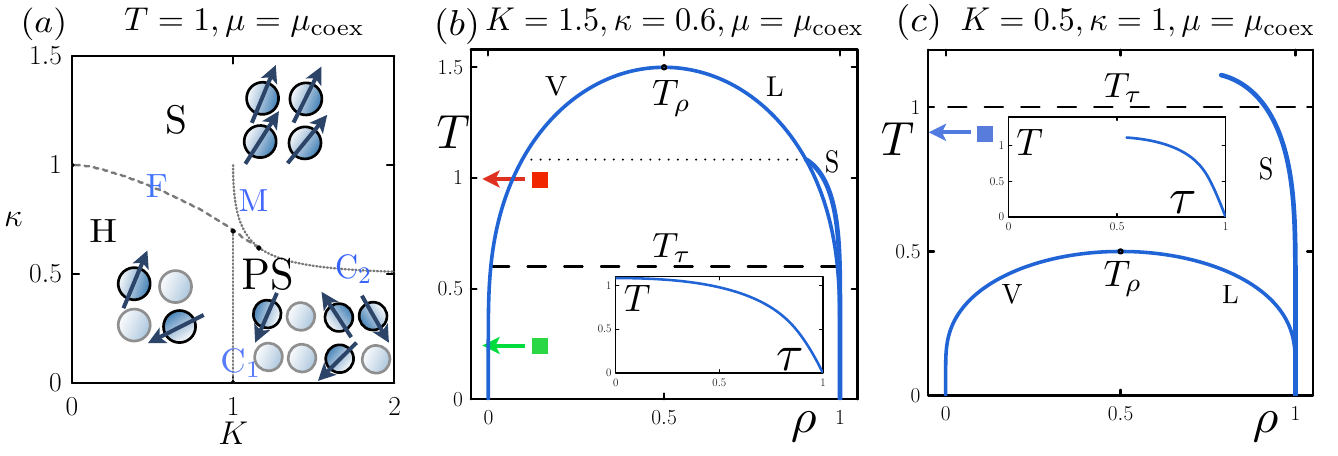} 
\caption{\label{fig1} Thermodynamic phase diagrams derived from \eqq{eq1}. (a) In the space of varying isotropic- $(K)$ and anisotropic $(\kappa)$ interaction strengths, we show regimes of stable homogeneous fluid H (moderate density, orientationally disordered); phase-separated (PS) liquid (L: high density, orientationally disordered) and vapor (V: low density, orientationally disordered); and solid S (high density, orientationally ordered). Cartoons depict the nature of these phases. Critical points and the nature of the lines F, M, C$_{1,2}$ are discussed in Section~\ref{sec_phase}. (b,c) Phase diagrams in the density ($\rho$)-temperature ($T$) plane for model parameters such that the solid phase emerges below (b) and above (c) the liquid-vapor critical point. We expect nonclassical ordering (when the solid is stable) for temperatures between the ordering temperatures $T_{\rho}$ and $T_{\tau}$ (marked). The insets to (b) and (c) show the emergence of solid order $\tau$ as a function of $T$ to be continuous and discontinuous, respectively. Langevin trajectories at the three marked temperatures (red, green and blue arrows) are shown in Fig.~\ref{fig2}.
}
 \end{figure*}

\begin{figure*}[] 
\label{}
\centering
\includegraphics[width=\linewidth]{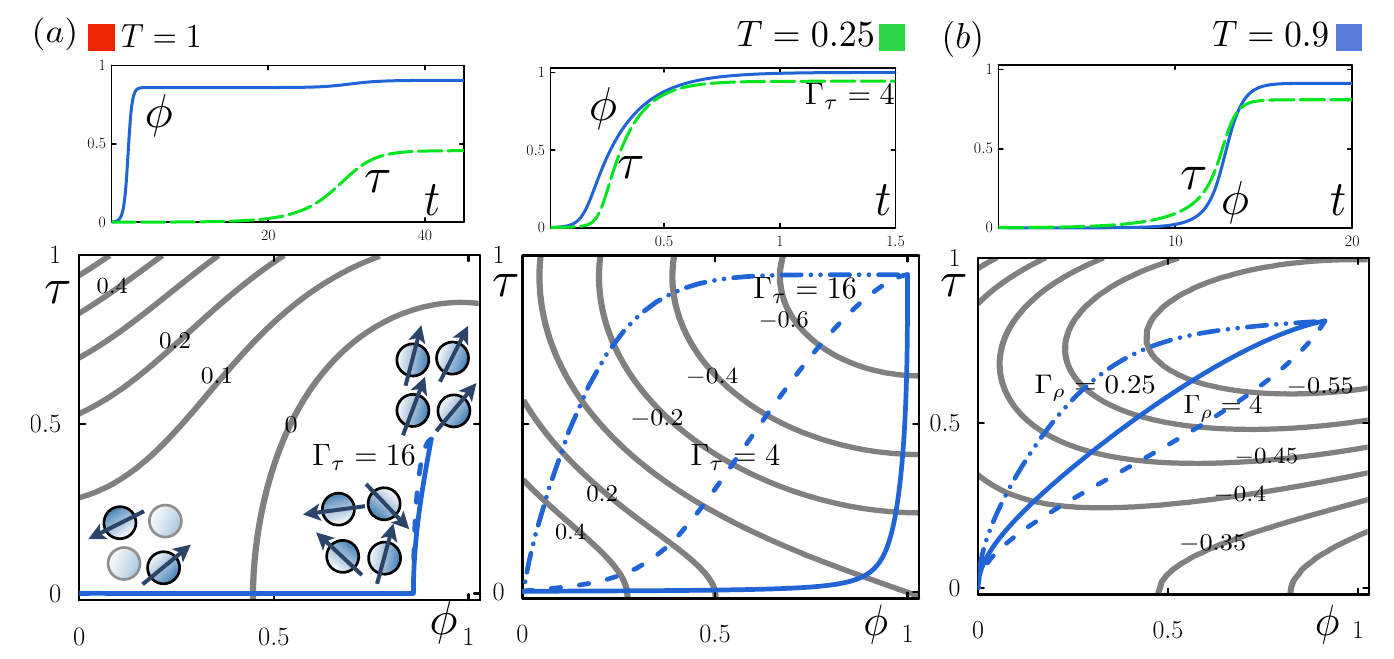} 
\caption{\label{fig2} Thermodynamically preferred assembly pathways derived from Langevin evolution on the free energy surface \eqq{eq1}, with initial conditions $(\phi(0),\tau(0))=(10^{-3},10^{-3})$. Order parameter mobilities $\Gamma_{\rho}$ and $\Gamma_{\tau}$ are set to unity unless otherwise marked. Top: order parameters versus time; bottom: assembly pathways (blue) plotted atop free energy contours (gray) with time as a parameter. Panels (a) show trajectories at two temperatures under conditions used to generate Fig.~\ref{fig1}(b). At the higher temperature ($T=1$), the nonclassical `density-structure' pathway is favored thermodynamically, because the fluid phase ($\phi=0=\tau$) is unstable to perturbations of density $\phi$ but not to perturbations of structure $\tau$. Trajectories generated using structural mobilities $\Gamma_{\tau}=1$ (solid blue line) and $\Gamma_{\tau}=16$ (dotted blue line) almost superpose. Cartoons depict the nature of three points along the trajectory. At the lower temperature ($T=0.25$), by contrast, the fluid phase is unstable in both directions in order parameter space, and no thermodynamic bias for nonclassical ordering exists. The trajectory followed depends on order parameter mobilities. (b) Assembly at $T=0.9$ under conditions used to generate Fig.~\ref{fig1}(c). Here the structure-density pathway is favored thermodynamically.}
 \end{figure*}

With the phase behavior of the model established, we turn to the question of how the solid phase emerges if it is stable and if the system is prepared in the homogeneous fluid phase H ($\phi=0=\tau$). We imagine this latter phase, which is of moderate density and possesses no orientational order, to describe a well-mixed system. The thermodynamic driving force associated with evolution of the bulk phase from H to the solid is connected to the stability of the free energy surface, in the vicinity of H, in the $\phi$- and $\tau$-directions of order parameter space. These stabilities can be assessed by Taylor expansion of \eqq{eq1}. Retaining only those terms required for thermodynamic stability (and ignoring field-independent terms) we find 
\bea
\label{expansion}
f_{\rm eff}(\rho,\tau) &\approx& \frac{1}{2}K \left(1 - \beta K \right)\phi^2 + 2 \k \left(1 -\beta \k \right) \tau^2 \nonumber \\
&+&c_{40} \phi^4+c_{06} \tau^6 -c_{12} \phi \tau^2+c_{14} \phi \tau^4 + c_{24} \phi^2 \tau^4 +c_{32} \phi^3 \tau^2.
\eea
Recall that $K \equiv J/4$, $\kappa \equiv Q/4$, and $\phi \equiv 2 \rho-1$. The coefficients $c_{nm} \equiv \left(n!m!\right)^{-1} \partial^n_{\phi} \partial^m_{\tau} f_{\rm eff} (\phi,\tau)|_{\phi,\tau=0}$ are positive constants (for $K$, $\kappa > 0$). The signs of the coefficients of the quadratic terms determine the stability of the fluid phase H. We see by inspection that the fluid is unstable to perturbations of density below a temperature $T_{\rho}=K$ (recall that $\beta \equiv 1/T$), and unstable to perturbations of structure $\tau$ below a temperature $T_{\tau}= \kappa$. While $T_{\rho}$ is the liquid-vapor critical temperature, $T_{\tau}$ is not in general equal to the temperature at which the solid becomes stable. Ordering temperatures for specified model parameters are labeled in Fig.~\ref{fig1}(b,c)~\footnote{In a space of fixed temperature $T=1$ and varying $K$ and $\kappa$, considered in Fig.~\ref{fig1}(a), the coefficients of the quadratic terms of \eqq{expansion} change sign on the lines $K=1$ and $\kappa=1$, respectively.}.

If the ordering temperatures $T_{\rho}$ and $T_{\tau}$ are different, and if the assembly temperature $T$ lies between them, then there exists a thermodynamic driving force along a preferred direction of order parameter space, or, in other words, a thermodynamic impetus for nonclassical ordering. We can visualize the thermodynamically preferred assembly pathway by assuming evolution of the order parameters according to the equations
\beq
\label{ev_rho}
\dot{ \rho}=-\Gamma_{\rho} \, \partial_{\rho} f_{\rm eff}(\rho,\tau)
\eeq
 and 
\beq
\label{ev_tau}
\dot{ \tau}=-\Gamma_{\tau} \, \partial_{\tau} f_{\rm eff}(\rho,\tau).
\eeq
We assume the order parameter mobilities $\Gamma_{\rho}$ and $\Gamma_{\tau}$ to be constant, and we imagine them to be directly related to particles' translational- and rotational diffusion constants, respectively. It is likely that these approximations hold best in the case of one-component molecular crystallization. In general, order parameter mobilities will depend on the order parameters themselves, particularly whenever slow dynamics is encountered. Such is the case, for example, in models of systems undergoing gelation~\cite{sciortino1993interference} or vitrification~\cite{whitelam2004geometrical}; in systems in which strong bonds are formed (e.g. in zeolite synthesis~\cite{jorge2005modeling}); and in binary mixtures that exhibit slow inter-species mixing~\cite{sanz2007evidence,peters2009competing}. The dynamics considered here neglects several other important features of real systems, such as the effects of spatial diffusion, interfaces, and of assembly-impairing kinetic traps. Interfaces confer a surface tension between bulk phases, and can render order parameter mobilities anisotropic. In future work we will assess the extent to which the effects of surfaces on assembly can be captured by a Ginzburg-Landau expansion of the model defined by \eqq{eq0}, and whether such expansions offer an alternative microscopic route to `phase field' models of crystallization (see e.g.~\cite{t—th2010polymorphism,gr‡n‡sy2002nucleation}). Here we focus on the simple dynamics of Eqs.~(\ref{ev_rho}) and~(\ref{ev_tau}). We argue that this dynamics reveals, importantly, the thermodynamic preference for time-dependent evolution of bulk order.

In Fig.~\ref{fig2}(a) we show Langevin pathways at $T=1$ and $T=0.25$ for model parameters of Fig.~\ref{fig1}(b). Interpreted literally, the classical notion of assembly describes an approximately straight line trajectory between start- and end points in a phase space of $(\phi,\tau)$. By contrast, at the higher temperature the nonclassical `density-structure' pathway is dominant, regardless of order parameter mobilities (pathways for $\Gamma_{\tau}=1$ and $\Gamma_{\tau}=16$ nearly superpose), because the fluid H is stable to perturbations of structure but not of density. At the lower temperature the fluid is unstable in both directions of order parameter space, and both classical- and nonclassical pathways can be taken, depending upon order parameter mobilities. The density-structure pathway, characteristic of certain proteins' crystallization, owes its existence to the liquid-vapor critical point, as in the case of isotropic interactions. In panel (b) we show preferred pathways at $T=0.9$ for the model parameters of Fig.~\ref{fig1}(c). Here the nonclassical `structure-density' pathway, characteristic of some melts~\cite{vekilov2005two}, is preferred, though rapid evolution of $\rho$ results in near-classical behavior. 

\section{Intermediate solid phases}
\label{sec_extension}
\begin{figure}[] 
\label{}
\centering
\includegraphics[width=0.6\linewidth]{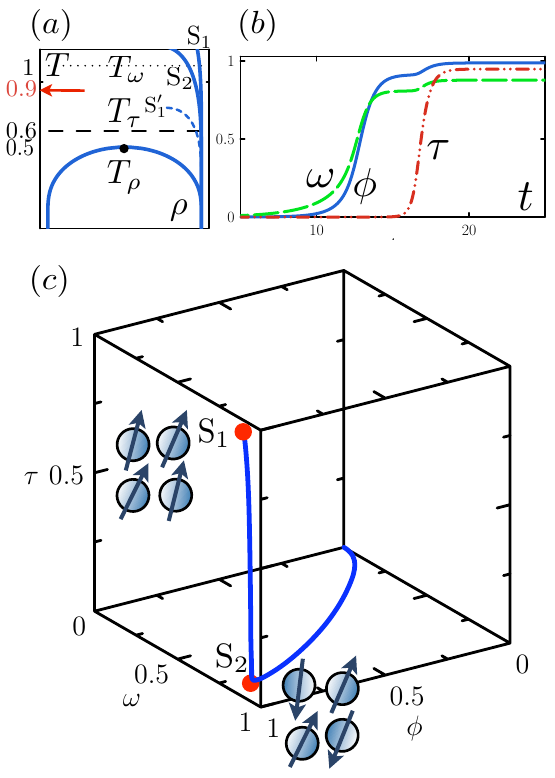} 
\caption{\label{fig3} Thermodynamics (a) and thermodynamically preferred assembly pathway (b,c) derived from \eqq{eq6}, for $K=0.5, \kappa=0.6,\kappa_2=1$. Panel (a) identifies two solid phases, a ferromagnetic phase S$_1$ and a nematic phase S$_2$, in addition to the liquid-vapor coexistence curve.  The assembly pathway shown in panels (b) (order parameters versus time) and (c) (parametric plot in order parameter space) is generated at $T=0.9$ by Langevin evolution on the free energy hypersurface \eqq{eq6}, starting from $(\phi(0),\tau(0),\omega(0))=10^{-3} (1,1,1)$, with equal order parameter mobilities. At this temperature the fluid phase H $(\phi=\tau=\omega=0)$ is unstable to perturbations of nematic structure $\omega$ but not to perturbations of ferromagnetic structure $\tau$ (because $T$ lies below $T_{\omega}$ and above $T_{\tau}$; see panel (a)). Assembly of the ferromagnetic phase therefore occurs via the nematic phase. Cartoons depict the nature of the solid phases. }
 \end{figure}

The density-structure pathway in our model is driven by the liquid-vapor critical point. However, recent work~\cite{galkin2000nucleation,liu2009self} suggests that crystallization can be induced by assembly of a dense phase possessing some of the symmetries of the crystal even {\em above} the liquid-vapor critical temperature. To rationalize such behavior within the framework discussed here we can add to \eqq{eq0} the nematic interaction term $\Delta U_{ij}=- Q_2 n_i n_j \cos \left(2 \theta_{ij} \right)$, where $\theta_{ij}$ is the angle between neighboring particle orientations. The effective dimensionless Helmholtz free energy density for this augmented model is
 \bea
 \label{eq6}
 f_{\rm eff}(\rho,\tau,\omega)& =&\frac{1}{2} \left(J \rho^2 + Q \tau^2 + Q_2 \omega^2\right) \nonumber \\
 &-& \kt \, \ln \left( 1 + e^{\beta(J \rho + \tilde{\mu})}  \cal{I}(\tau,\omega) \right),
 \eea     
 where $\cal{I}(\tau,\omega)\equiv\int_0^{2 \pi} d \theta \, e^{\beta Q \tau \cos \theta+\beta Q_2 \omega \cos \left(2 \theta\right)}.$ Here $\omega \equiv \langle n \cos \left(2 \theta\right) \rangle$ is a nematic order parameter. From this free energy we find, via Taylor expansion, the ordering temperature for $\omega$ to be $T_{\omega}=\kappa_2\equiv Q_2/4$. The phase diagram for $K=0.5,\kappa=0.6,\kappa_2=1$  is shown in Fig.~\ref{fig3}(a), labeled with the ordering temperatures $T_{\rho}$, $T_{\tau}$ and $T_{\omega}$; we focus on assembly at $T=0.9$ (arrow). Here we observe a stable ferromagnetic solid phase S$_1$ $(\phi_1,\omega_1,\tau_1)$=$(0.99,0.88,0.95)$ having free energy density $-1.1$, and an unstable nematic solid phase S$_2$ $(\phi_2,\omega_2,\tau_2)$=$(0.91,0.81,0)$ of free energy density $-0.6$. In the absence of the nematic coupling $\kappa_2$ the ferromagnetic solid (shown by line ${\rm S}_1'$ in (a)) is not viable at $T=0.9$. When $\kappa_2=1$ it becomes stable, but because $T$ lies above $T_{\tau}$ and below $T_{\omega}$ we observe (Fig~\ref{fig3}(b,c)) assembly of the ferromagnetic phase S$_1$ via the unstable nematic phase S$_2$, along the $\omega-\phi-\tau$ pathway. Thus, assembly via a dense intermediate phase, whose symmetries are partially commensurate with the stable solid, occurs well above the liquid-vapor critical temperature. While different in detail, this behavior echoes the notion of assembly via metastable ordered intermediates considered in Ref.~\cite{liu2009self}; here it occurs because the free energy structure local to the homogeneous fluid phase favors assembly of the unstable solid phase S$_2$, rather than its stable counterpart S$_1$.
 
\section{Conclusions}

We have used mean field theory to study two models prototypical of particles able to interact isotropically and anisotropically. While the approach considered here neglects important effects of surfaces, molecular detail and thermal fluctuations, it reveals that complex behavior can be driven by bulk free energy alone. We find that for a broad range of parameters the free energy structures of these models favor assembly of stable solid phases via intermediate phases, either amorphous or ordered. For other parameter choices, by contrast, assembly pathways are determined principally by dynamical considerations. One can observe in such cases classical pathways along which intermediate phases resemble the stable phase. The work presented here suggests a simple microscopic framework within which to rationalize and predict the assembly pathways of anisotropic particles.
 
\section{Acknowledgements} 

We thank Jim DeYoreo for discussions. This work was performed at the Molecular Foundry, Lawrence Berkeley National Laboratory, and was supported by the Director, Office of Science, Office of Basic Energy Sciences, of the U.S. Department of Energy under Contract No. DE-AC02--05CH11231 (75\% support) and as part of the Center for Nanoscale Control of Geologic CO$_2$, an Energy Frontier Research Center, under the same Contract No. (25\% support).

\break

\break

\section{Figures}

 \break

 \break

\end{document}